\documentclass[prl,twocolumn,showpacs,preprintnumbers,
  amsmath,amssymb,nofootinbib]{revtex4}
\usepackage{graphicx}
\usepackage{dcolumn}
\usepackage{bm}
\begin{document}
\newcommand{\be}{\begin{equation}}
\newcommand{\ee}{\end{equation}}
\newcommand{\lb}[1]{\label{#1}}
\newcommand{\ca}{{\mathcal A}}
\newcommand{\cu}{{\mathcal U}}
\newcommand{\cK}{{\mathcal K}}
\newcommand{\cT}{{\mathcal T}}
\newcommand{\ptl}{\partial}
\newcommand\lgth{[\,\text{\rm length}\,]}
\newcommand{\Sig}{\Sigma}
\newcommand{\Sigp}{\Sigma_{+}}
\newcommand{\Sigm}{\Sigma_{-}}
\newcommand{\Sigc}{\Sigma_{\times}}
\newcommand{\Udot}{\dot{U}}
\newcommand{\Nm}{N_{-}}
\newcommand{\Nc}{N_{\times}}
\newcommand{\enl}{\\\hfill\rule{0pt}{0pt}}
\newcommand{\sfrac}[2]{\frac{#1}{#2}}
\newcommand{\ct}[1]{\cite{#1}}
\newcommand{\mnote}[1]{\footnote{ MNOTE #1}}

\title{Asymptotic silence of generic cosmological singularities}

\author{Lars Andersson${}^{1}$}
\email{larsa@math.miami.edu}

\author{Henk van Elst${}^{2}$}
\email{H.van.Elst@qmul.ac.uk}

\author{Woei Chet Lim${}^{3}$}
\email{wclim@mathstat.dal.ca}

\author{Claes Uggla${}^{4}$}
\email{Claes.Uggla@kau.se}

\affiliation{${}^{1}\!$Department of Mathematics,
  University of Miami, Coral Gables, FL 33124, USA}

\affiliation{${}^{2}\!$Astronomy Unit, Queen Mary, University of
  London, Mile End Road, London E1 4NS, United Kingdom}

\affiliation{${}^{3}\!$Department of Mathematics and Statistics,
Dalhousie University, Halifax, Nova Scotia, Canada B3H 3J5}

\affiliation{${}^{4}\!$Department of Physics, University of
    Karlstad, S-651 88 Karlstad, Sweden}

\date{January 25, 2005}
\begin{abstract}
In this letter we investigate the nature of generic cosmological
singularities using the framework developed by Uggla {\em et al.\/}
We do so by studying the past asymptotic dynamics of general vacuum
$G_{2}$ cosmologies, models that are expected to capture the
singular behavior of generic cosmologies with no symmetries at
all. In particular, our results indicate that asymptotic silence
holds, i.e., that particle horizons along all timelines shrink to
zero for generic solutions. Moreover, we provide evidence that
spatial derivatives become dynamically insignificant along generic
timelines, and that the evolution into the past along such
timelines is governed by an asymptotic dynamical system which is
associated with an invariant set --- the silent boundary. We also
identify an attracting subset on the silent boundary that organizes
the oscillatory dynamics of generic timelines in the singular
regime. In addition, we discuss the dynamics associated with
recurring spike formation.

\end{abstract}
\maketitle
The singularity theorems of Penrose and Hawking~\cite{hawpen1970}
state that generic cosmological models contain an initial
singularity, but do not give any information on the nature of this
singularity. Heuristic investigations of this issue led
Belinski\v{\i}, Khalatnikov and Lifshitz~\cite{bkl1982} (BKL) to
propose that a generic cosmological initial singularity is
spacelike, local and oscillatory. Uggla {\em et
al.\/}~\cite{uggetal2003} (UEWE) reformulated Einstein's field
equations (EFEs) by introducing scale-invariant variables which
have the property that all individual terms in EFEs become
asymptotically bounded, for generic solutions. This made it
possible to characterize a generic cosmological initial singularity
in terms of specific limits. The numerical study of the picture
proposed by UEWE was initiated in Ref.~\cite{andetal2004},
specializing to Gowdy vacuum spacetimes which have a
non-oscillatory singularity. This letter presents the results of
the first detailed study of the oscillatory asymptotic dynamics of
inhomogeneous cosmologies from the dynamical systems point of view
introduced in UEWE.

Here we focus on vacuum cosmologies with an Abelian symmetry group
$G_{2}$ with two commuting spacelike Killing vector fields, and the
spatial topology of a 3-torus. This is arguably the simplest class
of inhomogeneous models that is expected to capture the properties
of a generic oscillatory singularity. Numerical investigations of
$G_{2}$ spacetimes supporting the BKL proposal were carried out by
Weaver {\em et al.\/} in Ref.~\cite{weaetal1998,beretal2001}.

UEWE used an orthonormal frame formalism and factored out the
expansion of a timelike reference congruence $\bm{e}_{0}$ by
normalizing the dynamical variables with the isotropic Hubble
expansion rate $H$ of $\bm{e}_{0}$. This yielded a dimensionless
state vector $\bm{X} = (E_{\alpha}{}^{i}) \oplus \bm{S}$, where
$E_{\alpha}{}^{i}$ are the Hubble-normalized components of the
spatial frame vectors orthogonal to $\bm{e}_{0}$; $\bm{e}_{\alpha}
= e_{\alpha}{}^{i}\,\ptl_{i}$, $E_{\alpha}{}^{i} =
e_{\alpha}{}^{i}/H$.

The approach to an initial singularity will be said to be {\em
asymptotically silent\/} for timelines along which
$E_{\alpha}{}^{i} \rightarrow 0$, and {\em asymptotically silent
and local\/} for timelines along which $E_{\alpha}{}^{i}
\rightarrow 0$ and $E_{\alpha}{}^{i}\,\ptl_{i}\bm{S} \rightarrow
0$; in the latter case $E_{\alpha}{}^{i} = 0$ defines an unphysical
invariant set, the silent boundary. [In UEWE the concept of a
``silent singularity" was defined. However, the possibility of
``recurring spike formation," discussed below, motivates the
present distinctions and definitions.] The evolution equations for
$\bm{S}$ on the silent boundary are identical to EFEs for spatially
self-similar (SSS) and spatially homogeneous (SH) models, in a
symmetry adapted Hubble-normalized orthonormal
frame~\cite{andetal2004}.

Motivated by the discussion in UEWE, we conjecture that $\cu_{\rm
vac}^{-}$, the union of the bounded vacuum SH Type--I (Kasner) and
SH Type--II subsets on the silent boundary, form an attracting
subset that organizes the oscillatory dynamics of generic timelines
approaching an asymptotically silent and local vacuum singularity.

To obtain the equations for vacuum $G_{2}$ cosmologies, we
introduce coordinates $\{t, x, y_{1}, y_{2}\}$ and an orthonormal
frame: $\bm{e}_{0} = N^{-1}\,\ptl_{t}$, $\bm{e}_{1} =
e_{1}{}^{1}\,\ptl_{x} + e_{1}{}^{2}\,\ptl_{y_{1}} +
e_{1}{}^{3}\,\ptl_{y_{2}}$, $\bm{e}_{2} =
e_{2}{}^{2}\,\ptl_{y_{1}}$ and $\bm{e}_{3} =
e_{3}{}^{2}\,\ptl_{y_{1}} + e_{3}{}^{3}\,\ptl_{y_{2}}$, cf.
Ref.~\cite{wai1981}; $N$ and $e_{\alpha}{}^{i}$ are functions of
$t$ and $x$ only. For comparison with previous work, $\bm{e}_{2}$
is aligned with the Killing vector field $\ptl_{y_{1}}$. We choose
$2\pi$-periodic coordinates $x$, $y_{1}$ and $y_{2}$, yielding a
spatial 3-torus topology, and a temporal gauge such that the area
density of the $G_{2}$-orbits is given by $\ca :=
(e_{2}{}^{2}\,e_{3}{}^{3})^{-1} \propto e^{-t}$; this is convenient
since the level sets of $\ca$ give a global foliation for maximally
globally hyperbolic vacuum $G_{2}$ cosmologies~\cite{beretal1997},
and since $t \to +\infty$ at the singularity~\cite{isewea2003}. The
$G_{2}$ symmetry implies $0 = \bm{e}_{2}(f) = \bm{e}_{3}(f)$ for
any coordinate scalar $f$. Thus, only $N^{-1}\,\ptl_{t}$ and
$e_{1}{}^{1}\,\ptl_{x}$ act nontrivially on coordinate scalars, and
hence the equations of all spatial frame variables except
$e_{1}{}^{1}$ decouple; the essential Hubble-normalized variables
are thus $E_{1}{}^{1} := e_{1}{}^{1}/H$ and a subset of connection
components, which depend on $t$ and $x$ only. Inserting the above
restrictions into the relations in App.~5 of UEWE yields $0 =
A^{\alpha} = N_{1\alpha} = N_{33} = \Sig_{12} = \Udot_{2} =
\Udot_{3}$ ($\alpha = 1, 2, 3$), and the spatial frame gauge $R_{1}
= -\Sig_{23}$, $R_{2} = -\Sig_{31}$, $R_{3} = 0$. In addition, it
is convenient to define: $\Sigp := \frac{1}{2}(\Sig_{22}+\Sig_{33})
= -\frac{1}{2}\Sig_{11}$, $\Sigm :=
\frac{1}{2\sqrt{3}}(\Sig_{22}-\Sig_{33})$, $\Sigc :=
\frac{1}{\sqrt{3}}\Sig_{23}$, $\Sig_{2} :=
\frac{1}{\sqrt{3}}\Sig_{31}$, $\Nm := \frac{1}{2\sqrt{3}}N_{22}$
and $\Nc := \frac{1}{\sqrt{3}}N_{23}$. The Hubble-normalized
variables have the following physical interpretation: $\Sigp$,
$\Sigm$, $\Sigc$, $\Sig_{2}$ are shear variables for $\bm{e}_{0}$;
$\Udot = \Udot_{1}$ describes the acceleration of $\bm{e}_{0}$;
$\Nm$ and $ \Nc$ are spatial connection components that determine
the spatial curvature; $R_{\alpha}$ yields the angular velocity of
the spatial frame $\{\,\bm{e}_{\alpha}\,\}$. The lapse function is
given by $N = -\frac{1}{2}H^{-1}(1+\Sigp)^{-1}$. The deceleration
parameter $q$ and the spatial Hubble gradient $r$ are defined by
$(q+1):= -\,H^{-1}\,\bm{\ptl}_{0}H$ and $r :=
-\,H^{-1}\,\bm{\ptl}_{1}H$, respectively, with $\bm{\ptl}_{0} :=
-2(1+\Sigp)\,\ptl_{t}$ and $\bm{\ptl}_{1} :=
E_{1}{}^{1}\,\ptl_{x}$. These definitions yield the {\em
integrability condition\/} $\bm{\ptl}_{0}r - \bm{\ptl}_{1}q =
(q+2\Sigp)\,r - (r-\Udot)\,(q+1)$.

Imposing the above restrictions and gauge choices on EFEs in vacuum
yields the following {\em evolution equations\/} and {\em
con\-straints\/}:
\begin{subequations}
\lb{eq:tlaevol}
\begin{align}
\lb{tlae11dot} \bm{\ptl}_{0}E_{1}{}^{1}
& = (q+2\Sigp)\,E_{1}{}^{1} \\
\lb{tlasigpdot} \bm{\ptl}_{0}(1+\Sigp)
& = (q-2)\,(1+\Sigp) + 3\Sig_{2}^{2} \\
\lb{tlasig2dot} \bm{\ptl}_{0}\Sig_{2}
& = (q-2-3\Sigp+\sqrt{3}\Sigm)\,\Sig_{2} \\
\lb{tlasigmdot} \bm{\ptl}_{0}\Sigm + \bm{\ptl}_{1}\Nc
& = (q-2)\,\Sigm + (r-\Udot)\,\Nc \nonumber \\
& + 2\sqrt{3}\Sigc^{2} - 2\sqrt{3}\Nm^{2}
- \sqrt{3}\Sig_{2}^{2} \\
\lb{tlancdot} \bm{\ptl}_{0}\Nc + \bm{\ptl}_{1}\Sigm
& = (q+2\Sigp)\,\Nc + (r-\Udot)\,\Sigm \\
\lb{tlasigcdot} \bm{\ptl}_{0}\Sigc - \bm{\ptl}_{1}\Nm
& = (q-2-2\sqrt{3}\Sigm)\,\Sigc \nonumber \\
& - (r-\Udot+2\sqrt{3}\Nc)\,\Nm \\
\lb{tlanmdot} \bm{\ptl}_{0}\Nm - \bm{\ptl}_{1}\Sigc
& = (q+2\Sigp+2\sqrt{3}\Sigm)\,\Nm \nonumber \\
& - (r-\Udot-2\sqrt{3}\Nc)\,\Sigc \ ,
\end{align}
\end{subequations}
and
\begin{subequations}
\label{eq:tlavacg2constr}
\begin{align}
\lb{tlarea}
0 & = (\bm{\ptl}_{1}-r+\Udot)\,(1+\Sigp) \\
\lb{tlagauss} 0 & = 1
- (\Sigp^{2}+\Sig_{2}^{2}+\Sigm^{2}+\Nc^{2}+\Sigc^{2}+\Nm^{2}) \\
\lb{tlacodac1}
0 & = (1+\Sigp)\,\Udot + 3(\Nc\,\Sigm-\Nm\,\Sigc) \\
\lb{tlacodac3} 0 & = (\bm{\ptl}_{1}-r+\sqrt{3}\Nc)\,\Sig_{2} \ ,
\end{align}
\end{subequations}
where $q := 2(\Sigp^{2}+\Sigm^{2}+\Sigc^{2}+\Sig_{2}^{2}) -
\sfrac{1}{3}\,(\bm{\ptl}_{1}-r+\Udot)\,\Udot$. Since we are
concerned with generic features, we restrict to the case $\Sigp
\neq -1$ ($\Sigp = -1$ yields the Minkowski spacetime). We use the
gauge constraint~(\ref{tlarea}) and the Codacci
constraint~(\ref{tlacodac1}) to solve for $r$ and $\Udot$ and so
obtain the reduced state vector $\bm{X} = (E_{1}{}^{1}, \Sigp,
\Sig_{2}, \Sigm, \Nc, \Sigc, \Nm) = (E_{1}{}^{1}) \oplus \bm{S}$.
Note that the Gau\ss\ constraint~(\ref{tlagauss}) implies that the
components of $\bm{S}$ are bounded.

Because of the symmetry restrictions, $E_{1}{}^{1}$ is the only
spatial frame variable in our state space; in the present context
asymptotic silence is thus associated with $E_{1}{}^{1} \rightarrow
0$, while $E_{1}{}^{1} = 0$ is referred to as the silent
boundary. Our numerical experiments, which employ the {\tt
RNPL}~\cite{marsa1995} and {\tt CLAWPACK}~\cite{lev1999} packages
with up to $2^{16}$ spatial grid points on the $x$-interval
$(0,2\pi)$, indicate that asymptotic silence holds in the present
$G_{2}$ case for {\em all\/} timelines of a generic solution.
Indeed, our numerical simulations indicate that ${\rm
max}_{x}(E_{1}{}^{1})$ decays exponentially. Moreover, they
indicate that $\lim_{t \to +\infty} (\Delta:= ||
E_{1}{}^{1}\,\ptl_{x}\bm{S} ||^{2}) = 0$ along generic timelines of
a generic solution, i.e., generically the singularity is
asymptotically silent and local, and hence in this case the
asymptotic dynamics is governed by the equations on the silent
boundary.

On the silent boundary $E_{1}{}^{1} = 0$, the integrability
condition and Eq.~(\ref{tlancdot}) yield $r^{2}= -3f\Nc^{2}$, while
Eq.~(\ref{tlacodac3}) reduces to $0 =
(r-\sqrt{3}\Nc)\,\Sig_{2}$. In contrast to Ref.~\cite{andetal2004},
we are here concerned with the general case $\Sig_{2} \neq 0$, for
which $r = \sqrt{3}\Nc$ and hence $f =-1$; in this case the
equations on the silent boundary are identical to the
Hubble-normalized equations of the exceptional SSS
Type--$_{-1}\!$VI$_{0}$ models; see Wu~\cite[p.~635]{wu1981}.

Our numerical experiments suggest that, in addition to $E_{1}{}^{1}
\to 0$, $\bm{C} := (\Udot, r, \Nc, \Nm\Sigc) \rightarrow 0$ holds
for generic timelines of a generic solution when $ t \to
+\infty$. {\em On\/} the silent boundary $E_{1}{}^{1} = 0$, $\bm{C}
= 0$ yields the Kasner and SH Type--II subsets which are defined by
$0 = E_{1}{}^{1} = \Nm = \Nc = \Udot = r$, $1 =
\Sigp^{2}+\Sigm^{2}+\Sigc^{2}+\Sig_{2}^{2}$, $q = 2$ and $0 =
E_{1}{}^{1} = \Sigc = \Nc = \Udot = r$, $q = 2(1-\Nm^{2})$,
respectively.

\begin{figure}[!tbp]
\begin{tabular}{cc}
\includegraphics[scale=0.60]{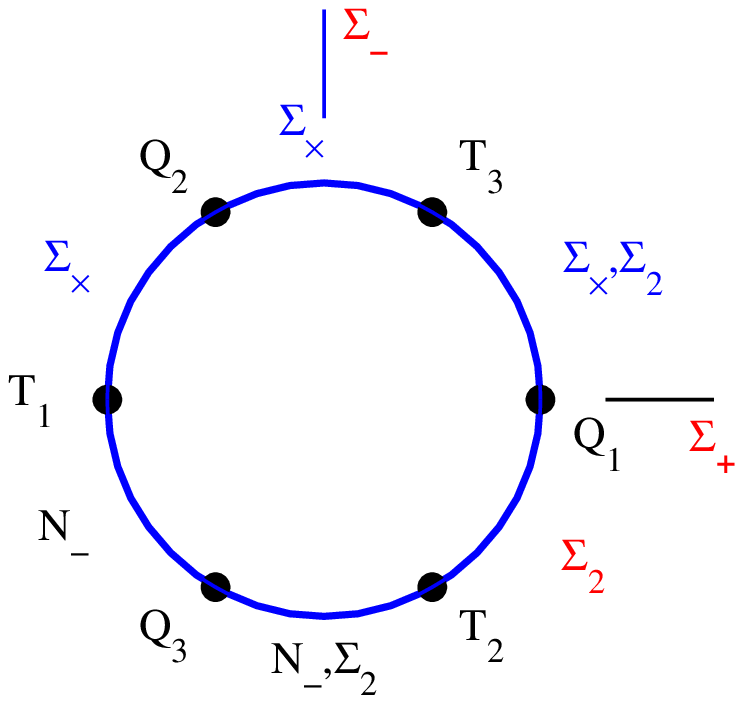}
\hspace*{.18in} &
\includegraphics[scale=0.60]{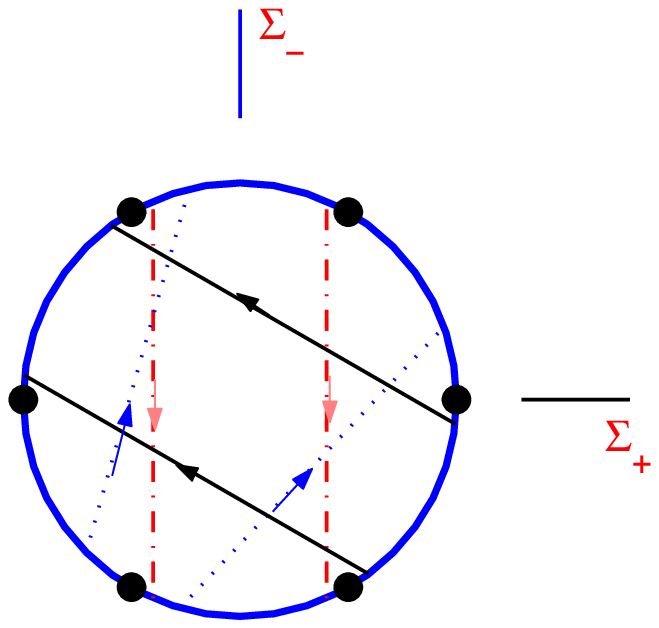}
\vspace*{.1in} \\
(a) & (b) \\
\end{tabular} 
\caption{(a) Unstable variables on $\cK$, and (b) single transition
  sets associated with $\Nm$ (dotted), $\Sigc$ (dash-dotted) and
  $\Sig_{2}$ (solid).}
\label{fig:kasner}
\end{figure}

With the present gauge choices, the Kasner subset contains a
subset of equilibrium points: $0 = E_{1}{}^{1} = \Sig_{2} = \Nc =
\Sigc = \Nm = \Udot = r$, $1 = \Sigp^{2} + \Sigm^{2}$, the Kasner
circle, $\cK$, which plays an essential role for the asymptotic
dynamics. A linear stability analysis of $\cK$ shows that all
variables are stable when $t \to +\infty$, except for $(\Nm,
\Sigc, \Sig_{2})$ which obey:
\begin{subequations}
\label{eq:unstable}
\begin{align}
\lb{nmunst}
\Nm & = \hat{N}_{-}\,e^{-[1-k(x)]t} \\
\lb{sigcunst}
\Sigc & = \hat{\Sig}_{\times}\,e^{-k(x)t} \\
\lb{sig2unst} \Sig_{2} & = \hat{\Sig}_{2}\,e^{[3-k(x)][1+k(x)]t/4}
\ ;
\end{align}
\end{subequations}
$E_{1}{}^{1}$ and $\bm{C}$ decay exponentially and uniformly
[$\Nm\Sigc \propto \exp(-t)$]. Here ``hatted'' variables are
functions of $x$ only. On $\cK$, $\Sigp=\hat{\Sig}_{+}$,
$\Sigm=\hat{\Sig}_{-}$, and $k(x) :=
-\sqrt{3}\hat{\Sig}_{-}(x)/[1+\hat{\Sig}_{+}(x)]$. Thus, $\Nm$,
$\Sigc$, $\Sig_{2}$ are unstable when $k(x) > 1$, $k(x) < 0$, $-1 <
k(x) < 3$, respectively; see Fig.~\ref{fig:kasner}(a).
The unstable mode $\Nm$ induces physical curvature transitions,
associated with the SH Type--II subset on $E_{1}{}^{1}=0$, while
$\Sigc$ and $\Sig_{2}$ induce frame transitions that lead to
rotations of the spatial frame and multiple representations of the
same solution, see Fig.~\ref{fig:kasner}(b); nevertheless, for the
present frame choice it is these gauge transitions that make
repeated curvature transitions possible, and hence they have
indirect physical implications. The $\Nm$, $\Sigc$ and $\Sig_{2}$
transitions imply that $k(x)$ changes according to the rules $k
\rightarrow 2-k$, $k \rightarrow -k$ and $k \rightarrow
(k+3)/(k-1)$, respectively.

The variables and equations that describe $\cu_{\rm vac}^{-}$ on
$E_{1}{}^{1}=0$ and the exceptional SH Type--VI$^{*}_{-1/9}$ case,
as given by Hewitt {\em et al.\/}~\cite{hewetal2003}, are
identical. As shown in Sec.~5 of Ref.~\cite{hewetal2003}, there
exist two integrals that describe the transition orbits. Although
multiple transitions are possible, single transitions increasingly
dominate. However, since frame transitions constitute gauge effects
we will not pursue this further. What is important physically is
that the variety $\cu_{\rm vac}^{-}$ induces an infinite sequence
of Kasner states related by SH Type--II curvature transitions
according to the frame invariant BKL map: $u \rightarrow u-1$, if
$u \geq 2$, and $1/(u-1)$, if $1 < u < 2$, where $u$ is defined
frame invariantly by $\det\Sig_{\alpha\beta} = \sfrac{1}{3}
\Sig_{\alpha}{}^{\beta}\Sig_{\beta}{}^{\gamma}
\Sig_{\gamma}{}^{\alpha} = 2 - 27u^{2}(1+u)^{2}/(1+u+u^{2})^{3}$.

\begin{figure}[!tbp]
\begin{tabular}{cc}
\includegraphics[scale=0.60]{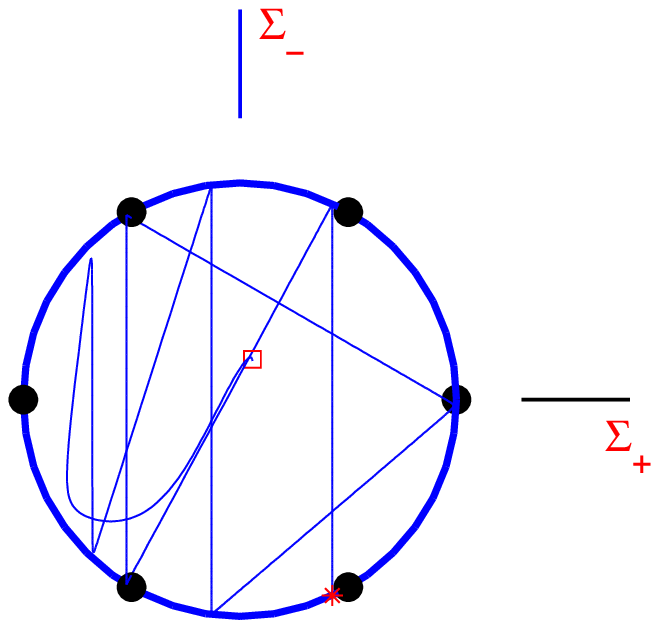}
\hspace*{.18in} &
\includegraphics[scale=0.60]{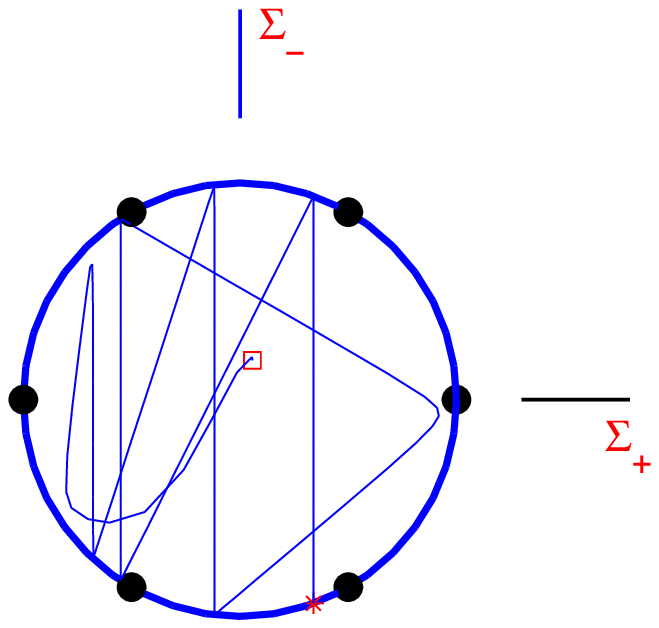} \\
(a) & (b)
\end{tabular}
\caption{Projections onto the $(\Sigp\Sigm)$-plane of a state space
orbit along the typical timeline $x = 0.3$ for (a) the full $G_{2}$
system, and (b) its restriction to the silent boundary. In both
cases the orbits approach $\cu_{\rm vac}^{-}$, i.e., $\Delta \to 0$
in (a), and $(E_{1}{}^{1}, \bm{C}) \to 0$ in both (a) and (b).}
\label{fig:timeline}
\end{figure}

Numerical investigations of vacuum SH Type--VI$^{*}_{-1/9}$ and SSS
Type--$_{-1}\!$VI$_{0}$ models indicate that generic solutions
asymptotically approach $\cu_{\rm vac}^{-}$. Our investigation
suggests that this is also true for the evolution associated with
generic timelines of the present inhomogeneous vacuum $G_{2}$
cosmologies, since our numerical results indicate that
$(E_{1}{}^{1},\bm{C},\Delta) \to 0$ when $t \to +\infty$ for a
generic timeline, see Fig.~\ref{fig:timeline}, and that thus the
BKL map holds asymptotically for such a timeline.

Belinski\v{\i}~\cite{bel1992} expressed concern that spatial
structure, created by the effect of different timelines going
through transitions at different times, could cause problems for
the BKL scenario. Numerical experiments show that this is not the
case for generic timelines. The reason is that spatial structure,
created by the mechanism described above, develops on superhorizon
scales; $\Delta \to 0$ within the shrinking particle horizons of a
generic timeline when $t \to +\infty$.

Our investigations indicate that $E_{1}{}^{1} \to 0$ as $t \to
+\infty$ for all timelines of a generic solution, and that
$(\bm{C},\Delta) \to 0$ for generic timelines so that
$\cu_{\rm vac}^{-}$ is a local past attractor. However, there are
indications that spiky features, closely related to the spikes in
Gowdy vacuum spacetimes, form along exceptional timelines; for such
timelines neither $\bm{C}$ nor $\Delta$ has a limit.

\begin{figure}[!tbp]
\centering
\includegraphics[scale=0.7]{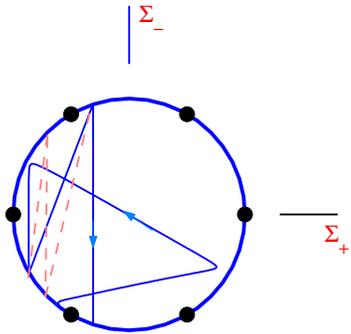}
\caption{Projection onto the $(\Sigp\Sigm)$-plane of a state space
  orbit undergoing a spike transition, followed by $\Sig_{2}$ and
  $\Sigc$ induced frame transitions and another spike transition
  (solid). The combination of $\Nm$ curvature and $\Sigc$ frame
  transitions corresponding to the second spike transition is shown
  (dashed). See also Ref.~\cite[p.~152]{wcl2004}.}
\label{fig:spiketrans}
\end{figure}

Recall that for the present general $G_{2}$ case, the whole of
$\cK$ is unstable with respect to at least one of $\Nm$, $\Sigc$,
$\Sig_{2}$; see Eqs.~(\ref{eq:unstable}). As in the Gowdy case,
spike formation is caused by the occurrence of a zero for one of
these variables at a point $(t, x(t))$ when the system is close to
$\cK$. It follows from Eqs. (\ref{tlasig2dot}) and
(\ref{tlacodac3}) that, for a generic smooth solution, $\Sig_{2}$
cannot cross zero and thus produces no spikes. Spikes in $\Sigc$
are ``false'' (gauge) spikes, while spikes in $\Nm$ are ``true''
(physical) effects which yield inherently inhomogeneous dynamical
features in the Hubble-normalized Weyl curvature scalars. Linear
analysis at $\cK$ shows
%
$E_{1}{}^{1}  \propto {\hat E_{1}{}^{1}}\,e^{-t} $ and 
%
%
$$
E_{1}{}^{1}\ptl_{x}\Nm  \propto \hat{E}_{1}{}^{1}
\left[\ptl_{x}\hat{N}_{-} + t\,\hat{N}_{-}\,\ptl_{x}k(x)\right]
e^{-[2-k(x)]t} \ .
$$
%
The state space orbits of the spatial points outside the particle
horizon of $(t,x(t))$, defined by $\Nm(t,x(t))=0$, undergo
curvature transitions with $\Nm = {\cal O}(1)$ and opposite signs
on either side of $x(t)$ when $k(x)>1$; since $x(t)$ does not go
through such a transition this leads to the formation of a
spike. For $k(x)>2$, $\bm{\ptl}_{1}\Nm = E_{1}{}^{1}{\ptl}_{x}\Nm $
is unstable on $\cK$, and hence grows in modulus at $(t,x(t))$,
which leads to a growth in modulus of $\Sigc$. Since the particle
horizon size at $(t,x(t))$ is of order $E_{1}{}^{1}$, see UEWE, the
above implies that $\ptl_{x} \Nm$ grows to order $1/E_{1}{}^{1}$
and that $\bm{\ptl}_{1} \Nm$, and thus also $\Delta$, is then
${\cal O}(1)$ at $(t,x(t))$. It therefore follows that the dynamics
fails to be local. Moreover, one can similarly argue that $\bm{C}$
becomes ${\cal O}(1)$. Since the dynamics fails to be local at
$(t,x(t))$, it is not governed by the silent boundary dynamical
system. Nevertheless, our investigations indicate that the
asymptotic dynamics is quite simple, and that it is related to that
on the silent boundary. Numerical simulations show that the orbit
described by $\bm{S}(t,x(t))$ during the formation and smoothing
out of a spike is described by the map $k \to 4-k$, equivalent to a
sequence of local $\Nm$--$\Sigc$--$\Nm$ transitions; following
Ref.~\cite{wcl2004}, we refer to this behavior as a {\em spike
transition\/}, see Fig. \ref{fig:spiketrans}.
The simplicity of this structure suggests that there may exist an
effective dynamical system governing the spike transitions, playing
a role analogous to that of the silent boundary system.

Numerical investigations suggest that an isolated zero in $\Nm$ may
persist as $t \to +\infty$; if true this yields an infinite
sequence of recurring spike transitions. Since the horizon scale
decays exponentially, one expects $x(t)$ to converge exponentially
to a point $x_{\rm spike}$, and since the dynamics fails to be
local during spike transitions, it follows that $\bm{C}$ and
$\Delta$ fail to have limits along the timeline $x=x_{\rm spike}$;
we refer to such a timeline as a {\em spike timeline\/}. Timelines
along which $(\bm{C},\Delta) \to 0$ as $t \to +\infty$ are called
{\em non-spike timelines\/}. Since the opportunities for new spike
formation occur at increasing time intervals due to the fact that
$\cK$ consists of equilibrium points for the system, and since the
horizon size decreases exponentially, we conjecture that generic
timelines are non-spike timelines. Our analysis supports the
conjecture that as $t \to +\infty$ the Kretschmann scalar becomes
unbounded, also along spike timelines.

For the Gowdy case ($\Sig_{2}=0$), the sequence of spike
transitions terminates when $k(x(t))$ reaches the interval $(0,2)$
on the lower part of $\cK$; see also Ref.~\cite{garwea2003}. If
$k(x(t))$ reaches the interval $(0,1)$, the spike disappears
completely, while in the interval $(1,2)$ a permanent spike is
formed for which $\Delta \to 0$. Hence, for the Gowdy case, $\Delta
\to 0$ uniformly, and thus the singularity is asymptotically silent
and local for {\em all\/} timelines.

The exponentially shrinking particle horizons of the spike
timelines cause severe numerical difficulties. Another obstacle are
the subsets associated with the Taub points $T_{\alpha}$
[cf. Fig.~\ref{fig:kasner}(a)]; in the present case $T_{1}$ in
particular. For SH Type--IX models, studied by
Ringstr\"om~\cite{rin2001}, the system spends a dominant portion of
its time undergoing oscillations in the vicinity of the Taub
points; this can be expected to hold also in the present $G_{2}$
case. In $G_{0}$ cosmologies, recently studied numerically by
Garfinkle~\cite{gar2003} in terms of the framework of UEWE, these
issues should cause formidable problems; their resolution is likely
to constitute a major step toward a rigorous analysis of generic
spacetime singularities, and an understanding of the cosmic
censorship problem.

We are grateful to Matt Choptuik for generous help in getting
started with the {\tt RNPL} package. We also thank John Wainwright
for many helpful discussions. LA is supported in part by the NSF,
contract no.~DMS 0407732. CU is supported by the Swedish Research
Council. LA thanks the Erwin Schr\"{o}dinger Institute, Vienna, and
the Albert Einstein Institute, Golm, and HvE and CU thank the
Department of Mathematics, University of Miami, for hospitality
during part of the work on this paper.

\bibliography{prlg2vac}

\end{document}